\documentclass[prl,twocolumn,showpacs,preprintnumbers,amsmath,amssymb,unsortedaddress,superscriptaddress]{revtex4-1}
\usepackage{graphicx}
\usepackage{dcolumn,color}
\usepackage{bm}

\usepackage{multirow}
\usepackage{epsfig}
\usepackage[english]{babel}

\begin{document}
\bibliographystyle{try}

\title{Well-established nucleon resonances revisited by double-polarization measurements}

\author{A.~Thiel}
\affiliation{Helmholtz--Institut f\"ur Strahlen-- und Kernphysik, Universit\"at Bonn, Germany}
\author{A.V.\hspace{0.5mm}Anisovich}
\author{D.\hspace{0.5mm}Bayadilov}
\affiliation{Helmholtz--Institut f\"ur Strahlen-- und Kernphysik, Universit\"at Bonn, Germany}
\affiliation{Petersburg Nuclear Physics Institute, Gatchina, Russia}
\author{B.\hspace{0.5mm}Bantes}
\affiliation{Physikalisches Institut, Universit\"at Bonn, Germany}
\author{R.\hspace{0.5mm}Beck}
\affiliation{Helmholtz--Institut f\"ur Strahlen-- und Kernphysik, Universit\"at Bonn, Germany}
\author{Yu.\hspace{0.5mm}Beloglazov}
\affiliation{Helmholtz--Institut f\"ur Strahlen-- und Kernphysik, Universit\"at Bonn, Germany}
\affiliation{Petersburg Nuclear Physics Institute, Gatchina, Russia}
\author{M.\hspace{0.5mm}Bichow}
\affiliation{Institut f\"ur Experimentalphysik I, Universit\"at Bochum, Germany}
\author{S.\hspace{0.5mm}B\"ose}
\author{K.-Th.\hspace{0.5mm}Brinkmann}
\affiliation{Helmholtz--Institut f\"ur Strahlen-- und Kernphysik, Universit\"at Bonn, Germany}
\affiliation{II. Physikalisches Institut, Universit\"at Gie{\ss}en, Germany}
\author{Th.\hspace{0.5mm}Challand}
\affiliation{Physikalisches Institut, Universit\"at Basel, Switzerland}
\author{V.\hspace{0.5mm}Crede}
\affiliation{Department of Physics, Florida State University, Tallahassee, USA}
\author{F.\hspace{0.5mm}Dietz}
\author{P.\hspace{0.5mm}Drexler}
\affiliation{II. Physikalisches Institut, Universit\"at Gie{\ss}en, Germany}
\author{H.\hspace{0.5mm}Dutz}
\author{H.\hspace{0.5mm}Eberhardt}
\author{D.\hspace{0.5mm}Elsner}
\author{R.\hspace{0.5mm}Ewald}
\author{K.\hspace{0.5mm}Fornet-Ponse}
\affiliation{Physikalisches Institut, Universit\"at Bonn, Germany}
\author{St.\hspace{0.5mm}Friedrich}
\affiliation{II. Physikalisches Institut, Universit\"at Gie{\ss}en, Germany}
\author{F.\hspace{0.5mm}Frommberger}
\affiliation{Physikalisches Institut, Universit\"at Bonn, Germany}
\author{Ch.\hspace{0.5mm}Funke}
\author{M.\hspace{0.5mm}Gottschall}
\author{M.\hspace{0.5mm}Gr\"uner}
\author{E.\hspace{0.5mm}Gutz}
\author{Ch.\hspace{0.5mm}Hammann}
\affiliation{Helmholtz--Institut f\"ur Strahlen-- und Kernphysik, Universit\"at Bonn, Germany}
\author{J.\hspace{0.5mm}Hannappel}
\affiliation{Physikalisches Institut, Universit\"at Bonn, Germany}
\author{J.\hspace{0.5mm}Hartmann}
\affiliation{Helmholtz--Institut f\"ur Strahlen-- und Kernphysik, Universit\"at Bonn, Germany}
\author{W.\hspace{0.5mm}Hillert}
\affiliation{Physikalisches Institut, Universit\"at Bonn, Germany}
\author{Ph. Hoffmeister}
\author{Ch.\hspace{0.5mm}Honisch}
\affiliation{Helmholtz--Institut f\"ur Strahlen-- und Kernphysik, Universit\"at Bonn, Germany}
\author{I.\hspace{0.5mm}Jaegle}
\affiliation{Physikalisches Institut, Universit\"at Basel, Switzerland}
\author{I.\hspace{0.5mm}J\"urgensen}
\author{D.\hspace{0.5mm}Kaiser}
\author{H.\hspace{0.5mm}Kalinowsky}
\author{F.\hspace{0.5mm}Kalischewski}
\affiliation{Helmholtz--Institut f\"ur Strahlen-- und Kernphysik, Universit\"at Bonn, Germany}
\author{ S.\hspace{0.5mm}Kammer}
\affiliation{Physikalisches Institut, Universit\"at Bonn, Germany}
\author{I.\hspace{0.5mm}Keshelashvili}
\affiliation{Physikalisches Institut, Universit\"at Basel, Switzerland}
\author{V.\hspace{0.5mm}Kleber}
\author{F.\hspace{0.5mm}Klein}
\affiliation{Physikalisches Institut, Universit\"at Bonn, Germany}
\author{E.\hspace{0.5mm}Klempt}
\affiliation{Helmholtz--Institut f\"ur Strahlen-- und Kernphysik, Universit\"at Bonn, Germany}
\author{B.\hspace{0.5mm}Krusche}
\affiliation{Physikalisches Institut, Universit\"at Basel, Switzerland}
\author{M.\hspace{0.5mm}Lang}
\affiliation{Helmholtz--Institut f\"ur Strahlen-- und Kernphysik, Universit\"at Bonn, Germany}
\author{I.\hspace{0.5mm}Lopatin}
\affiliation{Petersburg Nuclear Physics Institute, Gatchina, Russia}
\author{Y.\hspace{0.5mm}Maghrbi}
\affiliation{Physikalisches Institut, Universit\"at Basel, Switzerland}
\author{K.\hspace{0.5mm}Makonyi}
\author{V.\hspace{0.5mm}Metag}
\affiliation{II. Physikalisches Institut, Universit\"at Gie{\ss}en, Germany}
\author{W.\hspace{0.5mm}Meyer}
\affiliation{Institut f\"ur Experimentalphysik I, Universit\"at Bochum, Germany}
\author{J.\hspace{0.5mm}M\"uller}
\affiliation{Helmholtz--Institut f\"ur Strahlen-- und Kernphysik, Universit\"at Bonn, Germany}
\author{M.\hspace{0.5mm}Nanova}
\affiliation{II. Physikalisches Institut, Universit\"at Gie{\ss}en, Germany}
\author{V.\hspace{0.5mm}Nikonov}
\affiliation{Helmholtz--Institut f\"ur Strahlen-- und Kernphysik, Universit\"at Bonn, Germany}
\affiliation{Petersburg Nuclear Physics Institute, Gatchina, Russia}
\author{R.\hspace{0.5mm}Novotny}
\affiliation{II. Physikalisches Institut, Universit\"at Gie{\ss}en, Germany}
\author{D.\hspace{0.5mm}Piontek}
\affiliation{Helmholtz--Institut f\"ur Strahlen-- und Kernphysik, Universit\"at Bonn, Germany}
\author{G.\hspace{0.5mm}Reicherz}
\affiliation{Institut f\"ur Experimentalphysik I, Universit\"at Bochum, Germany}
\author{A.\hspace{0.5mm}Sarantsev}
\affiliation{Helmholtz--Institut f\"ur Strahlen-- und Kernphysik, Universit\"at Bonn, Germany}
\affiliation{Petersburg Nuclear Physics Institute, Gatchina, Russia}
\author{Ch.\hspace{0.5mm}Schmidt}
\affiliation{Helmholtz--Institut f\"ur Strahlen-- und Kernphysik, Universit\"at Bonn, Germany}
\author{H.\hspace{0.5mm}Schmieden}
\affiliation{Physikalisches Institut, Universit\"at Bonn, Germany}
\author{T.\hspace{0.5mm}Seifen}
\author{V.\hspace{0.5mm}Sokhoyan}
\affiliation{Helmholtz--Institut f\"ur Strahlen-- und Kernphysik, Universit\"at Bonn, Germany}
\author{V.\hspace{0.5mm}Sumachev}
\affiliation{Petersburg Nuclear Physics Institute, Gatchina, Russia}
\author{U.\hspace{0.5mm}Thoma}
\author{H.\hspace{0.5mm}van\hspace{0.5mm}Pee}
\author{D.\hspace{0.5mm}Walther}
\author{Ch.\hspace{0.5mm}Wendel}
\affiliation{Helmholtz--Institut f\"ur Strahlen-- und Kernphysik, Universit\"at Bonn, Germany}
\author{U.\hspace{0.5mm}Wiedner}
\affiliation{Institut f\"ur Experimentalphysik I, Universit\"at Bochum, Germany}
\author{A.\hspace{0.5mm}Wilson}
\affiliation{Department of Physics, Florida State University, Tallahassee, USA}
\author{A.\hspace{0.5mm}Winnebeck}
\author{Y.\hspace{0.5mm}Wunderlich}
\affiliation{Helmholtz--Institut f\"ur Strahlen-- und Kernphysik, Universit\"at Bonn, Germany}

\collaboration{The CBELSA/TAPS Collaboration}
\noaffiliation

\date{\today}

%-------------abstract----------------

\begin{abstract}
The first measurement is reported of the double-polarization
observable $G$ in photoproduction of neutral pions off protons,
covering the photon energy range from 620 to 1120\,MeV and the full
solid angle. $G$ describes the correlation between the photon
polarization plane and the scattering plane for protons polarized
along the direction of the incoming photon. The observable is highly
sensitive to contributions from baryon resonances. The new results
are compared to the predictions from SAID, MAID, and BnGa partial
wave analyses. In spite of the long-lasting efforts to understand
$\gamma p\to p\pi^0$ as the simplest photoproduction reaction,
surprisingly large differences between the new data and the latest
predictions are observed which are traced to different contributions
of the $N(1535)$ with spin-parity $J^P = 1/2^-$ and $N(1520)$ with
$J^P=3/2^-$.
 In the third resonance region, where $N(1680)$ with $J^P = 5/2^+$
production dominates, the new data are reasonably
close to the predictions.
\end{abstract}
\pacs{14.20}

%----------end of abstract-------------

\vskip 5mm

\maketitle

Symmetry arguments have led Gell-Mann and Zweig to introduce the
concept of quarks as constituents of mesons and baryons
\cite{GellMann:1962xb,Zweig:1964}. Later it was realized that quarks
carry a new kind of charge, color, and are the sources of gluons
transmitting the binding forces. A new theory evolved, quantum
chromodynamics \cite{Fritzsch:1975tx}, which proved to be very
successful for large momenta where the interaction becomes weaker.
In the region of meson and baryon resonances, QCD resists any
perturbative approach, and numerical calculations on a space-time
lattice are necessary. It has been only recently that the full
baryon spectrum including physical quantum numbers was presented
\cite{Edwards:2011jj}, even though unrealistically large quark
masses had to be used. Qualitatively, these lattice results confirm
the classic quark model calculations \cite{Isgur:1978}, whereas the
agreement with a fully relativistic model is less convincing
\cite{Loring:2001kx}. Yet the experimental mass spectrum exhibits
some remarkable differences to these calculations. The $N(1440)$
resonance with spin-parity $J^P = 1/2^+$ has a mass of 1440\,MeV and
falls below its spin-parity partner $N(1535)$ with $J^P = 1/2^-$
while QCD on the lattice and quark model calculations predict the
reverse. Also, above a mass of 1800\,MeV, the number of observed
resonances falls short of the number of predicted resonances.

\begin{figure*}[pt]
\begin{tabular}{cccc}
\raisebox{6mm}{\includegraphics[width=0.25\textwidth]{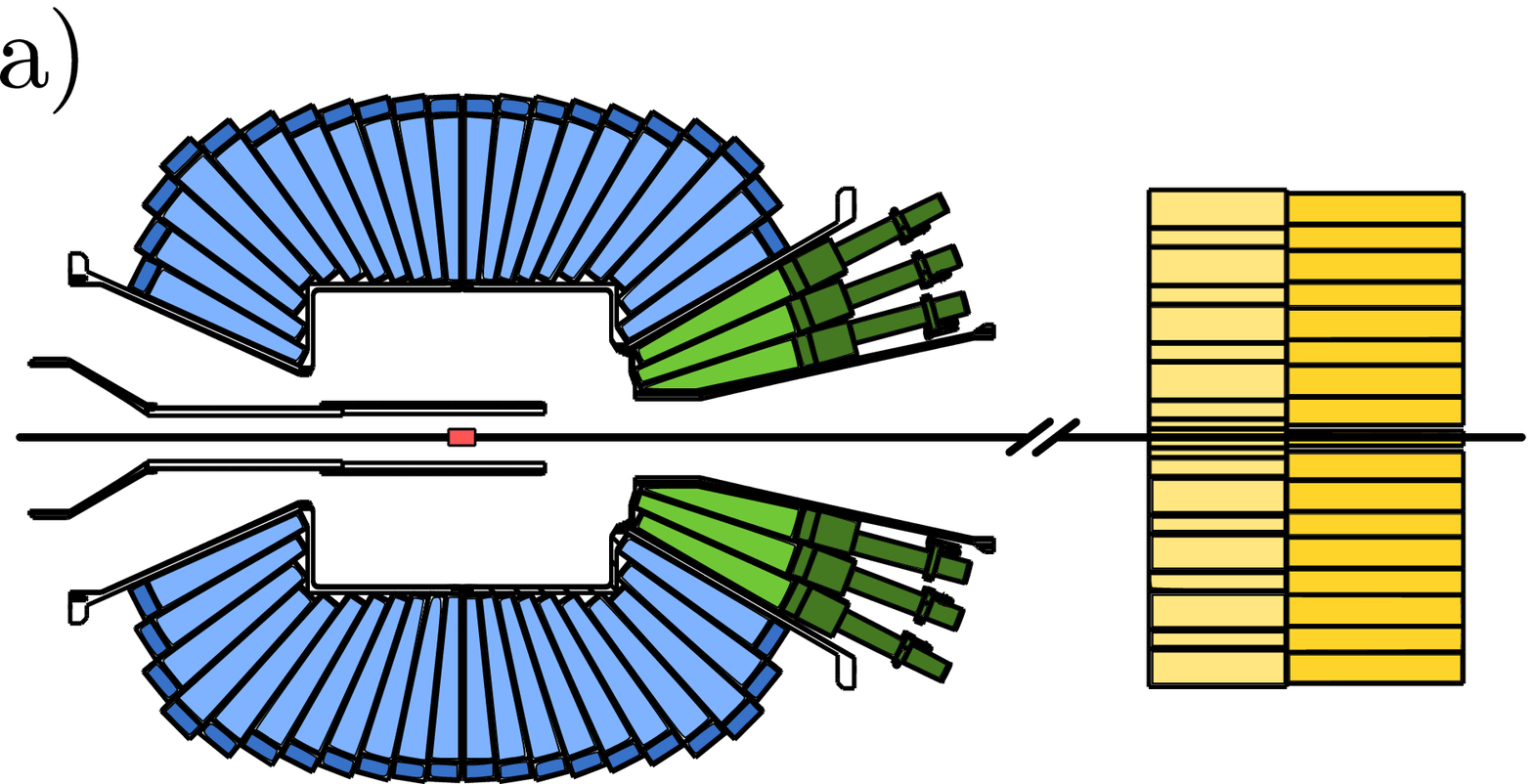}}&
\includegraphics[width=0.23\textwidth]{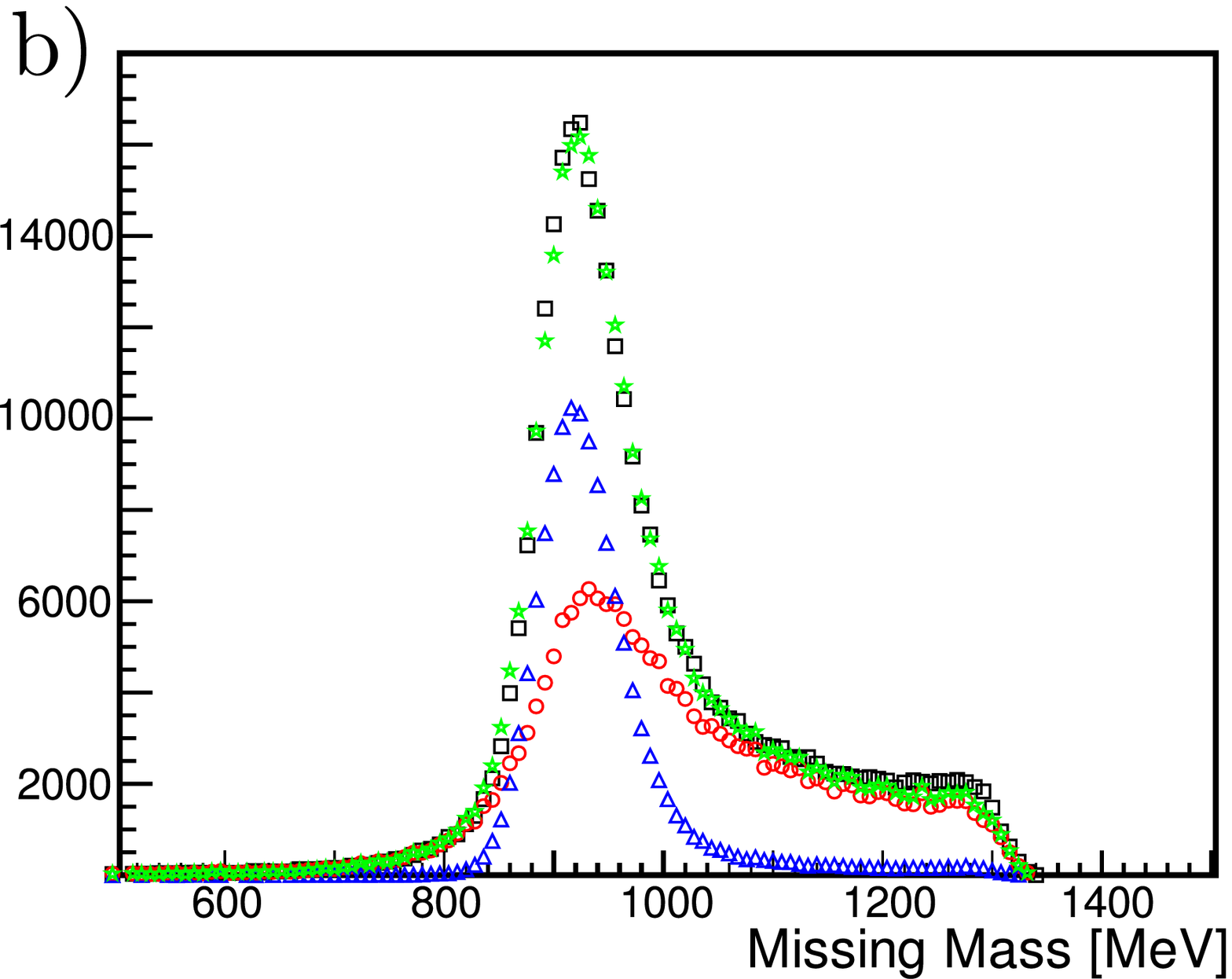}&
\includegraphics[width=0.23\textwidth]{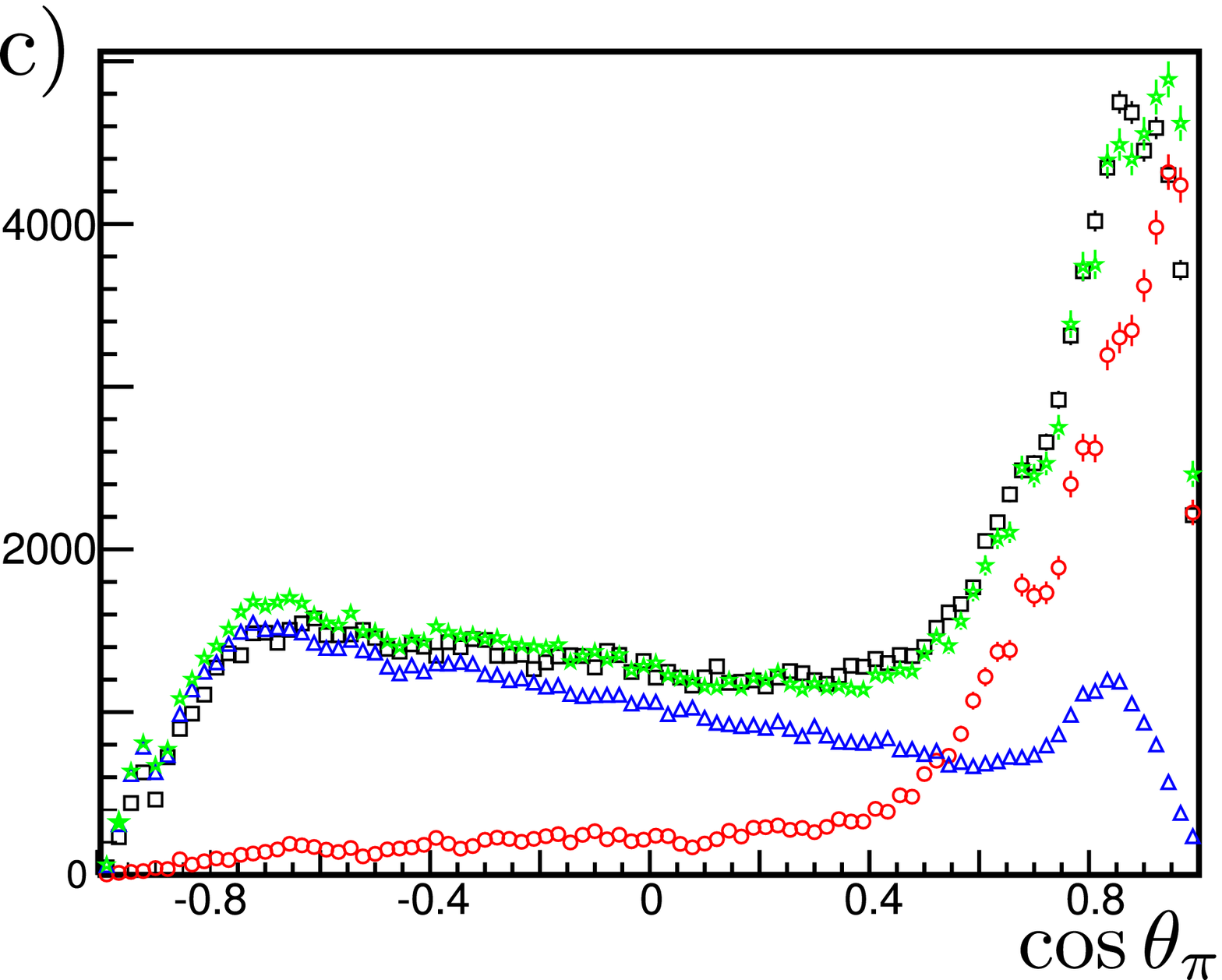}&
\includegraphics[width=0.23\textwidth]{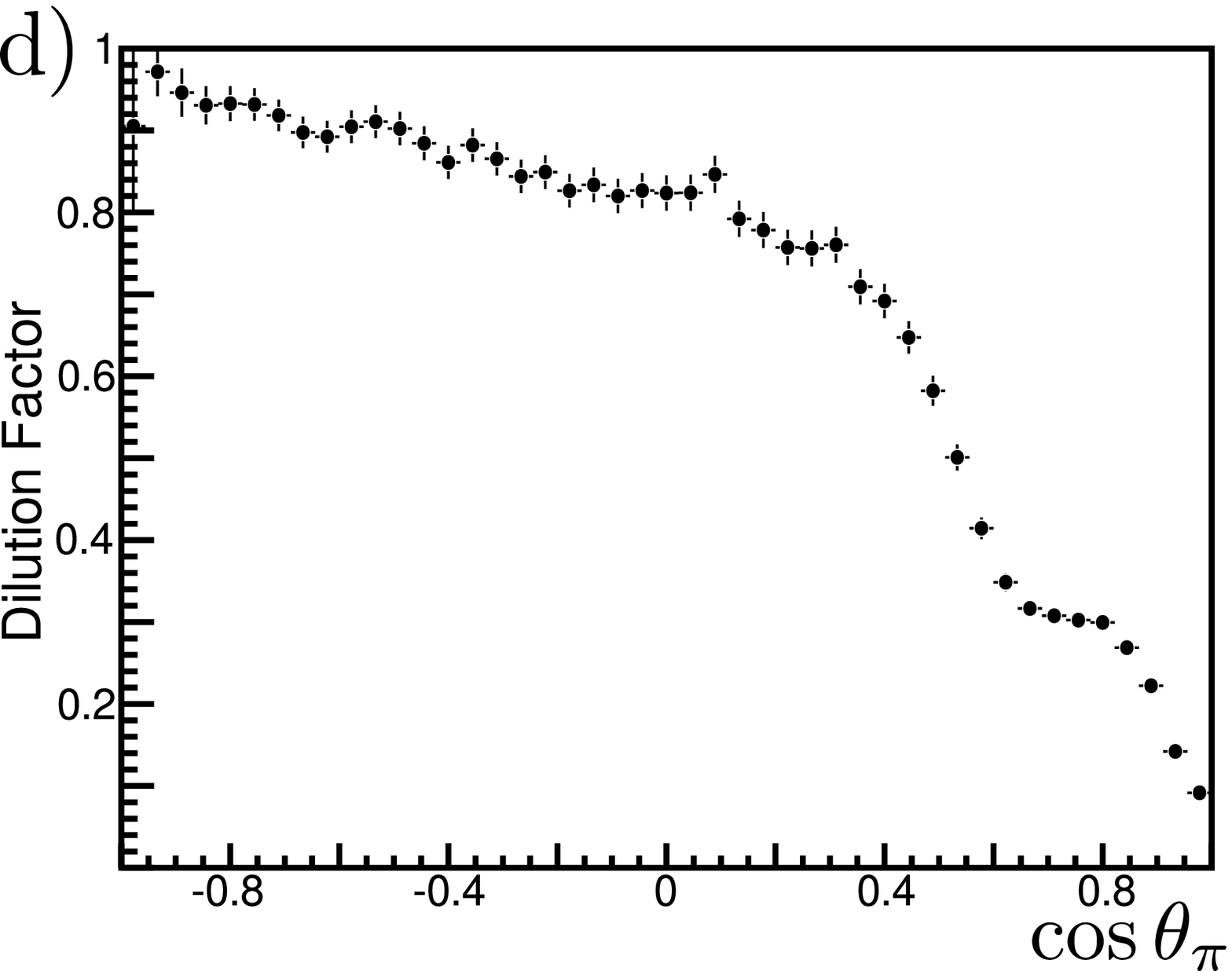}
\end{tabular}
\caption{\label{pic:dilution}(Color online) a) The central part of
the detector system. The CsI(Tl) crystals (blue and green) are read out via wavelength
shifters and photodiodes or photomultipliers, BaF$_2$ crystals
(yellow) in forward direction with photomultipliers. b) The missing
mass distribution, and (c) the $\pi^0$ angular distribution for
reaction (1) for an incident photon energy of $E_\gamma = 1000\pm25$~MeV;
butanol ({\scriptsize$\square$}), hydrogen
({\color{blue}$\triangle$}), carbon ({\color{red}$\circ$}), and the
sum of hydrogen and carbon data ({\color{green}$\ast$}). From these
distributions, the dilution factor (d) is determined.}
\end{figure*}

Baryon resonances have very short life times, hence many resonances
with different spin-parities overlap; their properties have to be
unfolded from data in partial wave analyses. Most partial wave
analyses are done using $\pi N$ elastic and charge exchange
scattering, but photoproduction experiments offer the opportunity of
analyzing resonances which were not accessible up to now. However,
the number of independent observables necessary for a unique partial
wave solution increases. Below the two-pion production threshold at
$E_\gamma=310$\,MeV, the Watson theorem \cite{Watson:1954} relates
the $\pi N$ phases to those in photoproduction of single pions. In
this case the measurement of the differential cross section
$d\sigma$ and of the photon beam asymmetry $\Sigma$ is sufficient to
determine precisely the s- and p-wave multipoles ($l_\pi = 0, 1$),
as well as the small ratio of electric quadrupole (E2) and magnetic
dipole (M1) amplitudes in the $N \rightarrow \Delta(1232)$
transition \cite{Beck:1997ew,Blanpied:1997zz}. Above the two-pion
threshold, but still in the mass range where nucleon resonances
carry only one unit of intrinsic orbital angular momentum, five
experiments are sufficient to construct the spin-helicity structure
of the photoproduction amplitude
\cite{Omelaenko1981,Wunderlich2012}. In general, the observation of
eight polarization observables with high precision is required to
construct the electric ($E$) and magnetic ($M$) multipoles governing
the process \cite{Chiang:1996em,Sandorfi:2010uv}.

\begin{figure*}[t]
\begin{tabular}{cc}
\includegraphics[width=0.31\textwidth]{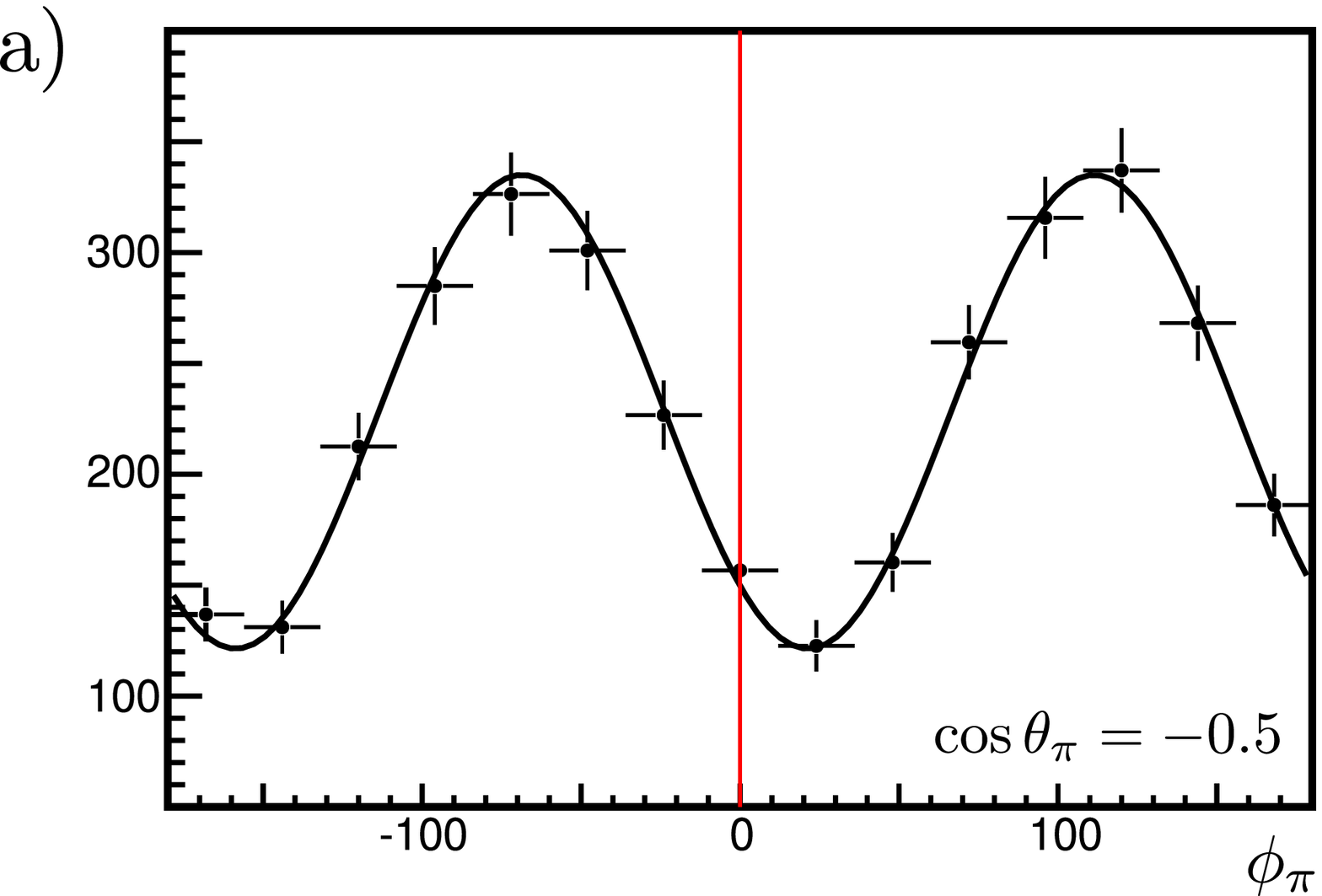}
\includegraphics[width=0.31\textwidth]{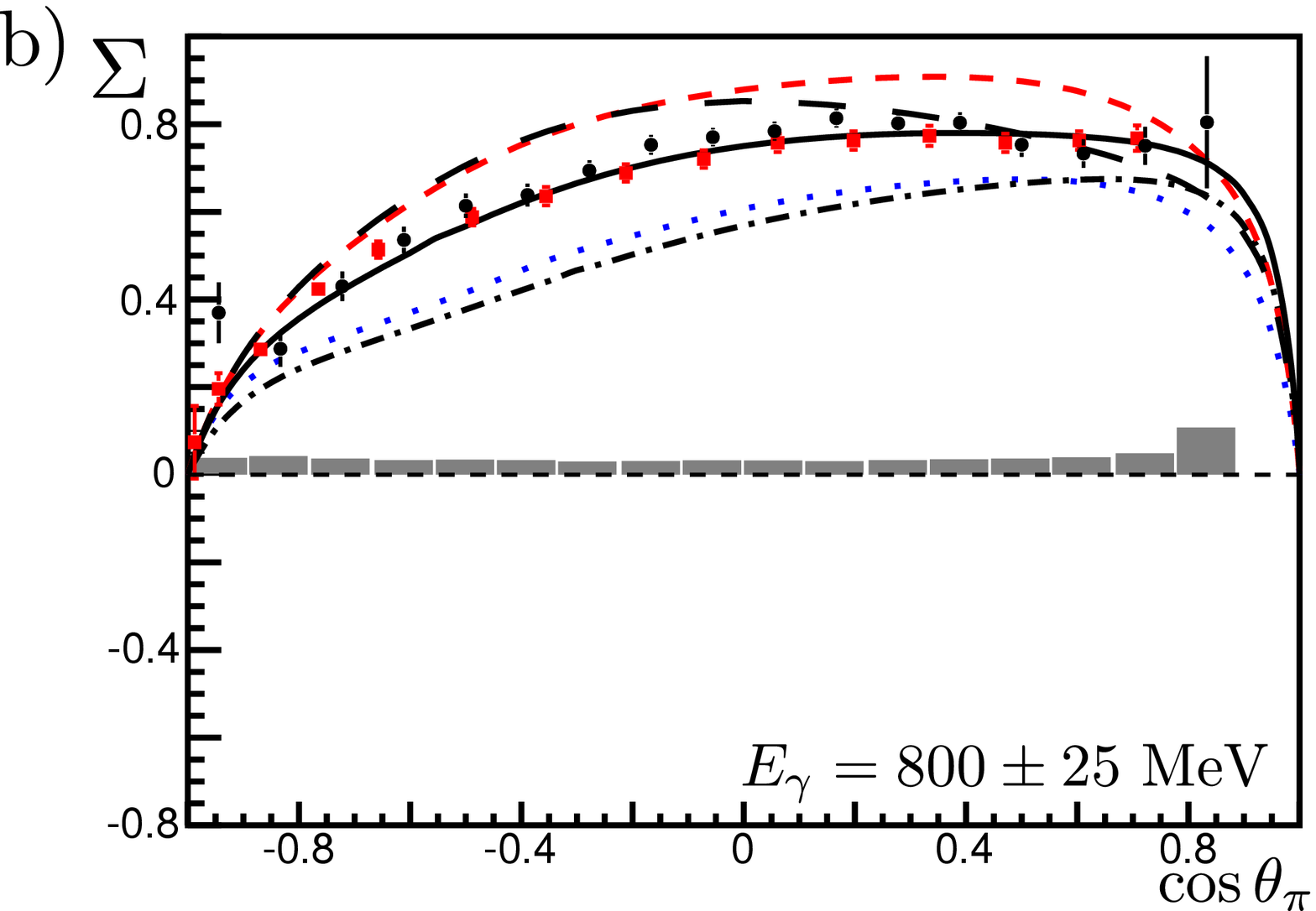}&
\includegraphics[width=0.31\textwidth]{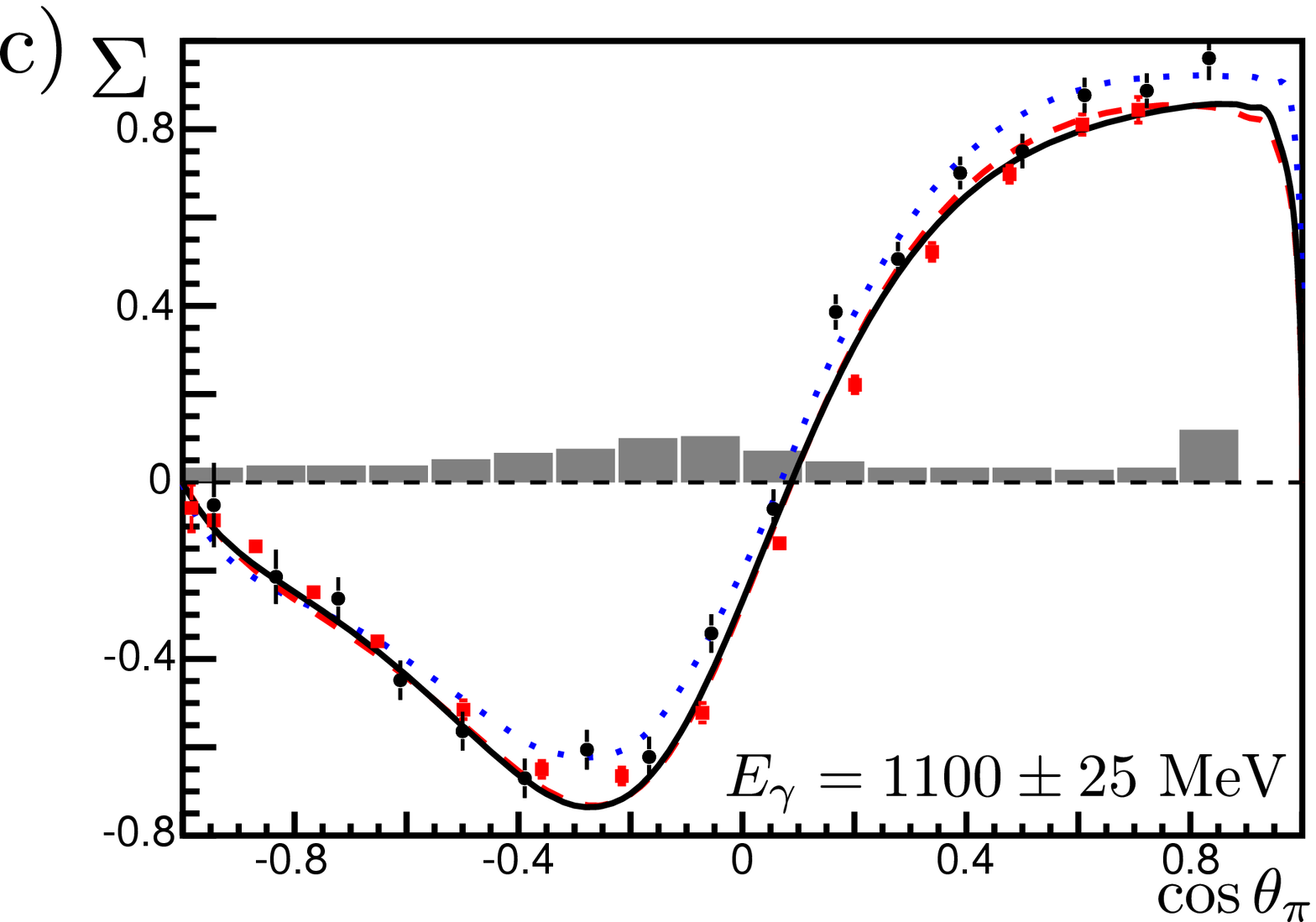}\vspace{-2mm}
\end{tabular}
\caption{\label{pic:Sigma}(Color online) a) A typical
$\phi_\pi$-distribution with a fit using eq. (2). b,c) The beam
asymmetry $\Sigma$ as a function of $\cos\theta_{\pi}$ for
$E_\gamma =800$\,MeV and for $E_\gamma =1100$\,MeV. Black dots show
our data, red squares GRAAL data \cite{Bartalini:2005wx}. The curves
represent predictions from different partial wave analyses. Solid
(black) curve: BnGa \cite{Anisovich:2012}; dashed (red): SAID
\cite{Dugger:2009pn}; long-dashed (black): BnGa with $E_{0^+}$ and
$E_{2^-}$ amplitudes from SAID; dotted (blue): MAID
\cite{Drechsel:1998hk}; dashed-dotted (black): BnGa with $E_{0^+}$
and $E_{2^-}$ amplitudes from MAID. Gray area shows the systematic error due
to interactions on nuclei and uncertainty in the photon polarization.}
\end{figure*}
In this letter we report on a first measurement of the double-polarization observable $G$ in photoproduction of neutral pions off
protons covering the second and third resonance region. Linearly polarized photons were produced by scattering of a 3.2\,GeV electron beam
\cite{Hillert:2006yb} off a diamond crystal \cite{Elsner:2008sn}, whereby maximal polarization of 65\% at 950\,MeV and
59\% at 1150\,MeV for a second data set were reached.
The photons then hit a butanol (${\rm C_4H_{10}O}$) target with longitudinally polarized protons~\cite{Dutz:2004zz}, with a mean proton
polarization of about 75\%. The butanol target was replaced by a
hydrogen or carbon target for background studies and for
normalization. The incoming photons may produce a $\pi^0$ in the
reaction
\begin{equation}
\vec\gamma\vec p\to p\pi^0.
\label{pinull}
\end{equation}

For linearly polarized photons (with polarization ${\rm p_\gamma}$)
and protons (with polarization ${\rm p_T}$), the number of events
$N$ at the polar angle $\theta_\pi$ due to reaction (\ref{pinull})
as a function of the azimuthal angle $\phi_\pi$ can be written in
the form
\begin{eqnarray}
\hspace{-3mm}\frac{N(\phi_\pi,\theta_\pi)}{N_{0}(\theta_\pi)} =1-
 {\rm p_\gamma} \Sigma_{\rm B} \cos(2\phi_\pi) +{\rm p_\gamma p_T} G_{\rm B}
\sin(2\phi_\pi)
\label{pol-h}
\end{eqnarray}
where $N_0$ is given by averaging $N(\phi_\pi,\theta_\pi)$ over $\phi_\pi$. The beam asymmetry
$\Sigma_{\rm B}$ and the observable $G_{\rm B}$ for the butanol target are related to the
corresponding quantities for scattering off free ($f$) protons
$\Sigma, G$ and bound ($b$) nucleons by
\begin{eqnarray}
\Sigma_{\rm B} = \frac{N_{0}^f\;\Sigma +
N_{0}^{b}\;\Sigma_{b}}{N_{0}^{f}+N_{0}^{b}};\quad G_{\rm B}
=\frac{N_{0}^f} {N_{0}^{f}+N_{0}^{b}} \cdot G.
\label{eqn:strahlasymmetrie}
\end{eqnarray}
The number of events in the denominator is the sum of the number of events where scattering
took place off protons or a bound nucleon, $N_0=
N_{0}^{f}+N_{0}^{b}$. There is no contribution of bound nucleons to
$G_{\rm B}$ since carbon and oxygen nuclei carry no polarization.
The quantity $D= N_{0}^f/(N_{0}^{f}+N_{0}^{b})$ is called dilution factor
and depends on the reaction and on the kinematical cuts performed. It is
also possible to determine $G$ without using the dilution
factor by reversing the target polarization and regarding the difference
$N^+(\phi_\pi,\theta_\pi)-N^-(\phi_\pi,\theta_\pi)$, given by $2
N_{0}^f {\rm p_\gamma p_T} G \sin(2\phi_\pi)$. A similar equation can be
written for a change of the photon polarization plane by $\pm\pi/4$ \cite{Thiel2012}.

Fig.~\ref{pic:dilution}a shows the calorimeter setup. Neutral pions are
reconstructed from their decay into two photons and a
measurement of their energy and direction in CsI(Tl) and
BaF$_2$ crystals enclosing the target hermetically.
The proton direction is determined, assuming that it
originated in the target center, from its hit in the surrounding
three-layer scintillation fiber detector
\cite{Suft:2005cq} or scintillation detectors in front of the
crystals in forward direction, and its hit in the CsI or
BaF$_2$ crystals. In comparison to \cite{Crede:2011dc}, two major
instrumental changes have been introduced. The forward opening of the
calorimeter is now covered by 90 additional CsI(Tl) crystals with
photomultiplier readout in the main calorimeter and 216 BaF$_2$ crystals,
2.1\,m downstream of the target. The tagger is now mounted
horizontally. Beam electrons not producing bremsstrahlung
are now deflected to the left and are stopped in a beam dump well
behind the detector system.

In the first step of the reconstruction, events due to reaction (1)
are selected using a series of kinematical cuts. Two classes of
events are retained, events with three hits in the calorimeters and
events with two hits. Both hits of a two-hit event are assumed to be
photons; their invariant mass is calculated and required to fall
into a $3\sigma$ ($\pm25$\,MeV) window centered at the $\pi^0$ mass.
The proton is assumed to be stopped in some material, but due to
energy and momentum conservation, the proton momentum can be
reconstructed.  If there is a third hit, one pair of hits is assumed
to be photons, and it is checked that the pair forms a $\pi^0$,
using again a $\pm3\sigma$ window. The third hit has to match the
direction of the proton momentum reconstructed as missing momentum.
Matching is defined by a cone, $\pm7^\circ$ in the azimuthal and
polar angles. Finally, a time coincidence is required between the
tagger hit and hits in one of the scintillation detectors or forward
calorimeter crystals. Additionally all events are removed, if the
calculated beam photon energy (under the assumption that the
reaction occurred on a proton) falls below the tagging threshold.
With these cuts the reconstruction efficiency for reaction
(\ref{pinull}) on free protons in the butanol target exceeds 60\%.

Fig. \ref{pic:dilution}b shows the resulting distribution of
missing masses for events passing the selection criteria, with
butanol, carbon, or hydrogen as target material. By proper scaling,
the sum of distributions from hydrogen and carbon reproduces very
well the distribution obtained with the butanol target. The missing
mass distributions show a peak at the proton mass and a tail due to
events where an additional meson was
produced but escaped detection. A final cut is now applied on the
missing mass at $m_p\pm50$\,MeV. The resulting $\cos \theta_{\pi}$
distribution of these events is exhibited in fig.
\ref{pic:dilution}c. Again, the sum of distributions from hydrogen
and carbon match the distribution obtained with the butanol target;
there is no new scaling factor involved.

For the double-polarization observable $G$ the correction factor for reactions on nuclei
 is the dilution factor, displayed in fig.
\ref{pic:dilution}d. Due to the kinematic cuts, the dilution factor
depends on both the photon energy $E_\gamma$ and the $\pi^0$
scattering angle $\theta_\pi$. For a backward $\pi^0$, the proton
momentum is high and therefore detected in the crystals with a high
efficiency.

A typical distribution of the measured $\phi_\pi$-dependent data yields with a fit using eq.
(2) is shown in fig. \ref{pic:Sigma}a.
Obviously, there is a large $\cos2\phi_\pi$ contribution but the
data are not symmetric with respect to $\phi_\pi=0$: there is a
significant $\sin2\phi_\pi$ contribution from which the desired
observable $G$ is deduced. Fig. \ref{pic:Sigma}b,c show the beam
asymmetry $\Sigma$ extracted from this and other similar plots for
two energy bins. The systematic error due to reactions on nuclei
are shown as gray bars. The new $\Sigma$ data are in excellent agreement with
earlier measurements of GRAAL \cite{Bartalini:2005wx}.

\begin{figure}[pt]
\includegraphics[width=0.46\textwidth]{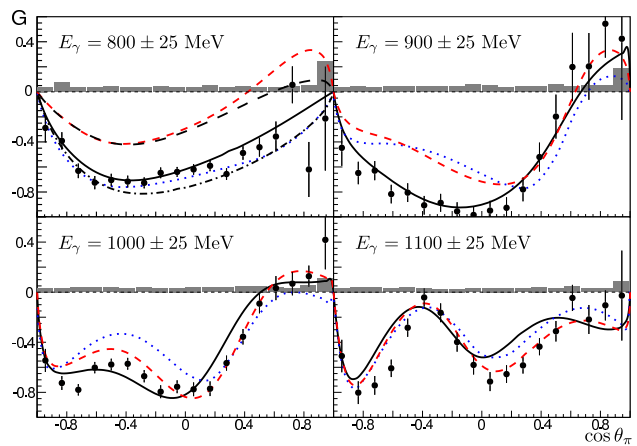}
\caption{\label{pic:results}(Color online) The polarization
observable $G$ as a function of $\cos\theta_{\pi}$ from $E_\gamma
=800$\,MeV up to $E_\gamma =1100$\,MeV. Systematic errors are shown
in gray bars. Curves: see fig. \protect\ref{pic:Sigma}.}
\end{figure}

The observable $G$ is determined from eq.
(\ref{pol-h}), as well as by reversing the target polarization, or
changing the photon polarization plane by $\pm \pi/4$. The results
are consistent, their spread is used to determine the systematic
errors. These include the systematic uncertainty in the dilution factor,
errors due to background in $\cos\theta_\pi > 0.9$, the
relative uncertainty in the target polarization of $\pm 2$\%, and the
relative uncertainty in the photon polarization of $\pm 5$\% \cite{Thiel2012b}.
Fig. \ref{pic:results} shows the angular distribution
of the polarization observable $G$ for selected photon energy ranges.
The error bars give the statistical errors, the systematic errors are
shown as gray bars.

The new data on $\Sigma$ and $G$ are compared to the predictions from
the SAID (SN11) \cite{Dugger:2009pn}, MAID \cite{Drechsel:1998hk}, and BnGa
\cite{Anisovich:2012} partial wave analyses. The results are
surprising.

\begin{figure}[pt]
\includegraphics[width=0.42\textwidth]{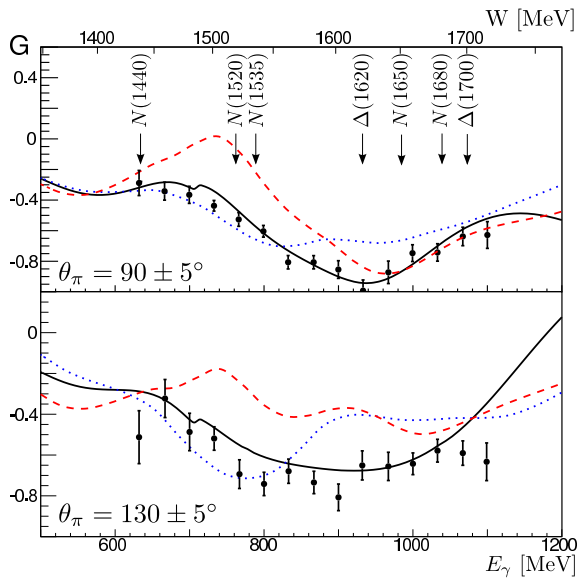}
\caption{\label{pic:angle}The double-polarization observable $G$ as
a function of energy for two selected bins in $\theta_\pi$. Curves:
see fig. \protect\ref{pic:Sigma}. For comparison the positions of
several resonances are marked.}
\end{figure}

Already at rather low energies, in the region of the four-star
resonances $N(1440)$, $N(1535)$ and $N(1520)$, very significant
differences in the predictions can be observed (see
figs.~\ref{pic:Sigma}~and~\ref{pic:results}). At $800$ and
$900$~MeV, the SAID prediction is incompatible with the data on
$\Sigma$ and $G$, MAID disagrees with the data on $\Sigma$ and on
$G$ at $900$~MeV. At $1000$ and $1100$\,MeV, BnGa, MAID and SAID
give a reasonable description, even though MAID is higher at
$1000$~MeV. These conclusions are supported by fig.~\ref{pic:angle}
where $G$ is shown as a function of $E_\gamma$ for two scattering
angles and compared to PWA predictions. At low energies, SAID
deviates strongly from the data; in the $900 - 1000$\,MeV range, the
consistency between data and MAID or SAID predictions is poor.
Above, none of the partial wave analyses reproduces the data.

We have studied possible reasons for the discrepant results. In this
discussion, the two electric multipoles $E_{0^+}$ and $E_{2^-}$ play
a leading role. If we replace the BnGa $E_{0^+}$ and $E_{2^-}$
multipoles by the corresponding SAID multipoles, we reproduce the
SAID results for $\Sigma$ and $G$ over a wide energy range, and the
same observation holds true for MAID. The reason for the deviation
between data and MAID and SAID is obviously traced to these two
multipoles. All other multipoles can be exchanged without a large
impact.

Resonances with no angular momentum between proton and $\pi^0$
($l_\pi=0$) contribute to the $E_{0^+}$ multipole which is sensitive
to the $N(1535)$, $\Delta(1620)$ and $N(1650)$ resonances. The
resonances $N(1520)$ and $\Delta(1700)$ with $l_\pi=2$ and total
spin $J=3/2$ contribute to the $E_{2^-}$ multipole. Indeed, the
three partial wave analyses BnGa, MAID, and SAID give significantly
different helicity amplitudes for these resonances, in particular
for the $A_{1/2}$ amplitude of $N(1535)$ and for $A_{3/2}$ of
$N(1520)$ (see Table~\ref{ph_coupl}).

In the medium energy range, in the third resonance region, the
$\gamma p\to p\pi^0$ cross section is dominated by the $J^P=5/2^+$
$N(1680)$ resonance. In this regime,~there~is good agreement between
the three predictions. The properties of this resonance are
obviously well defined: For the leading $A_{3/2}$ helicity
amplitude, BnGa finds $135\pm6$, MAID 134, and SAID $141\pm3$ (in
GeV$^{-\frac12}\,10^{-3}$).

In the high energy region, increasing differences between the predictions
can be observed for some angles. This is the so-called fourth resonance region where many resonances exist
with questionable evidence for their existence and, if they exist
at all, with at least poorly defined properties. New data on $G$ in this energy region
will help to disentangle the spectrum of nucleon resonances \cite{Thiel2012b}.
\begin{table}[t]
\begin{scriptsize}
\begin{center}
\renewcommand{\arraystretch}{1.5}
\hspace{-2mm}\begin{tabular}{|c|c|c|c|c|} \hline \boldmath$E_{0^+}$
&\multicolumn{1}{c|}{$N(1535)1/2^-$}
 &\multicolumn{1}{c|}{$N(1650)1/2^-$}
 &\multicolumn{1}{c|}{$\Delta(1620)1/2^-$}&\\
\hline Solution &$A^{1/2}$
         &$A^{1/2}$
         &$A^{1/2}$ & \\
\hline
BG2011-02&$105\pm 10$&$33\pm 7$&$52\pm 5$& \\
MAID-2007&$66$         &$33$         &$66$       &  \\
SAID-2011&$99\pm 2$&$65\pm 25$&$64\pm 2$& \\
\hline \hline \boldmath$E_{2^-}$
&\multicolumn{2}{c|}{$N(1520)3/2^-$}
 &\multicolumn{2}{c|}{$\Delta(1700)3/2^-$}\\
\hline Solution &$A^{1/2}$ &$A^{3/2}$
         &$A^{1/2}$ &$A^{3/2}$ \\
\hline
BG2011-02&$-22\pm 4$&$131\pm 10$&$160\pm 20$&\quad$165\pm 25$\quad\\
MAID-2007&$-27$         &$161$         &$226$         &\quad$210$\quad         \\
SAID-2011&$-16\pm 2$&$156\pm 2$&$109\pm 4$&\quad$84\pm 2$\quad\\
\hline
\end{tabular}\vspace{-5mm}
\end{center}\end{scriptsize} \caption{\label{ph_coupl}Helicity amplitudes of
low-lying negative-parity $N$ and $\Delta$ resonances contributing
to the $E_{0^+}$ and $E_{2^-}$ multipoles (in
GeV$^{-\frac12}\,10^{-3}$).}
\end{table}

Summarizing, we have reported the first measurement of the
double-polarization observable $G$ in the reaction $\vec{\gamma}
\vec{p} \rightarrow p \pi^0$ in a wide range of energies and
covering the full solid angle. The new data on $G$ resolve the
discrepant results on the helicity amplitudes of low-lying four-star
nucleon resonances obtained from BnGa, MAID, and SAID and are an
important step towards a complete data base which will define
unambiguously the nucleon excitation spectrum.
 \vspace{1mm}

We thank the technical staff of ELSA and the par\-ti\-ci\-pating
institutions for their invaluable contributions to the success of
the experiment. We acknowledge support from the \textit{Deutsche
Forschungsgemeinschaft} (SFB/TR16) and \textit{Schweizerischer
Nationalfonds}.

\end{document}